\input harvmac
\input epsf

\newcount\figno
\figno=0
\def\fig#1#2#3{
\par\begingroup\parindent=0pt\leftskip=1cm\rightskip=1cm\parindent=0pt
\baselineskip=12pt
\global\advance\figno by 1
\midinsert
\epsfxsize=#3
\centerline{\epsfbox{#2}}
\vskip 14pt

{\bf Fig. \the\figno:} #1\par
\endinsert\endgroup\par
}
\def\figlabel#1{\xdef#1{\the\figno}}
\def\encadremath#1{\vbox{\hrule\hbox{\vrule\kern8pt\vbox{\kern8pt
\hbox{$\displaystyle #1$}\kern8pt}
\kern8pt\vrule}\hrule}}

\overfullrule=0pt

\noblackbox
\parskip=1.5mm

\def\Title#1#2{\rightline{#1}\ifx\answ\bigans\nopagenumbers\pageno0
\else\pageno1\vskip.5in\fi \centerline{\titlefont #2}\vskip .3in}

\font\caps=cmcsc10

\noblackbox
\parskip=1.5mm

  
\def\npb#1#2#3{{\it Nucl. Phys.} {\bf B#1} (#2) #3 }
\def\plb#1#2#3{{\it Phys. Lett.} {\bf B#1} (#2) #3 }
\def\prd#1#2#3{{\it Phys. Rev. } {\bf D#1} (#2) #3 }
\def\prl#1#2#3{{\it Phys. Rev. Lett.} {\bf #1} (#2) #3 }
\def\mpla#1#2#3{{\it Mod. Phys. Lett.} {\bf A#1} (#2) #3 }

\def\cmp#1#2#3{{\it Commun. Math. Phys.} {\bf #1} (#2) #3 }

\def\bb#1{{\tt hep-th/#1}}

\def\jhep#1#2#3{{\it J. High Energy Phys.} {\bf #1} (#2) #3 }


           \def\CO{{\cal O}} 
   \def\CG{{\cal G}}
   
 \def\CR{{\cal R}}  
\def\CM{{\cal M}}  \def\CK{{\cal K}}
\def\CN{{\cal N}}


\def\dj{\hbox{d\kern-0.347em \vrule width 0.3em height 1.252ex depth
-1.21ex \kern 0.051em}}

\def\ket{\rangle}
\def\bra{\langle}

\def\Dirac{\,\raise.15ex\hbox{/}\mkern-13.5mu D}
\def\dirac{\,\raise.15ex\hbox{/}\kern-.57em \partial}
\def\shalf{{\ifinner {\textstyle {1 \over 2}}\else {1 \over 2} \fi}} 
\def\sshalf{{\ifinner {\scriptstyle {1 \over 2}}\else {1 \over 2} \fi}} 
\def\sfourth{{\ifinner {\textstyle {1 \over 4}}\else {1 \over 4} \fi}}

\lref\rmarti{T. Banks, M. R. Douglas, G. T. Horowitz and E. Martinec,
\bb{9808016.}} 

\lref\raw{J.J. Atick and E. Witten, \npb{310}{1988}{291.}}

\lref\rusex{J.L.F. Barb\'on and E. Rabinovici, \npb{545}{1999}{371} 
\bb{9805143.}}

\lref\rgutperlem{J.R. David, M. Gutperle, M. Headrick and S. Minwalla,
\bb{0111212.}}

\lref\rsen{A. Sen, \jhep{9806}{1998}{007}  \bb{9803194.}}

\lref\rbvm{J.L.F. Barb\'on and M.A. V\'azquez-Mozo, \npb{497}{1997}{236}  
\bb{9701142.}}
  
\lref\raps{A. Adams, J. Polchinski and E. Silverstein, \jhep{0110}{2001}{029}
  \bb{0108075.}}

\lref\rvafa{A. Dabholkar, \bb{0109019.} C. Vafa, \bb{0111051.} A. Dabholkar
and C. Vafa, \bb{0111155.} Y. Michishita and P. Yi, \bb{0111199.} }  

\lref\rtipocero{N. Seiberg and E. Witten, \npb{276}{1986}{272.}} 

\lref\rgaberdiel{O. Bergman and M.R. Gaberdiel, \jhep{9907}{1999}{022}
\bb{9906055.}} 

\lref\rrohm{R. Rohm, \npb{237}{1984}{533.}}

\lref\rcosta{M.S. Costa and M. Gutperle, \jhep{0103}{2001}{027} \bb{0012072.}}

\lref\rben{I. Affleck and A.W. Ludwig, \prl{67}{1991}{161.}}

\lref\rrelevance{J.A. Harvey, D. Kutasov and E.J. Martinec, 
\bb{0003101.}}

\lref\rraben{ S. Elitzur, E. Rabinovici, G. Sarkisian,  
\npb{541}{1999}{246}  
\bb{9807161.}}  

\lref\rmarinho{D. Kutasov, M. Mari\~no and G.W. Moore, \jhep{0010}{2000}
{045} \bb{0009148.}} 

\lref\rshatas{A.A. Gerasimov and S.L. Shatashvili, \jhep{0010}{2000}{034} 
\bb{0009103.}}

\lref\rdabho{A. Dabholkar, \bb{0111004.}}

\lref\rchicago{J.A. Harvey, D. Kutasov, E.J. Martinec and G. Moore,
\bb{0111154.}}

\lref\refr{S. Elitzur, A. Forge and E. Rabinovici,  
\npb{359}{1991}{581.}}

\lref\rabelr{S.A. Abel, J.L.F. Barb\'on, I.I. Kogan and
 E. Rabinovici, \bb{9911004.}} 

\lref\rliouville{S.R. Das, S. Naik and S.R. Wadia, {\it Mod. Phys. Lett.}
{\bf A4} (1989) 1033\semi
J. Polchinski, \npb{324}{1989}{123.}}

\lref\rgutstr{M. Gutperle and A. Strominger, \jhep{0106}{2001}{035} 
\bb{0104136.}} 

\lref\rhag{R. Hagedorn, {\it Supp. Nuovo Cim.} {\bf 3} (1965) 147.} 

\lref\rhist{S. Fubini and G. Veneziano, {\it Nuovo Cim.} {\bf 64A} (1969)
1640. K. Huang and S. Weinberg, \prd{25}{1970}{895.}
 S. Frautschi, \prd{3}{1971}{2821.} R.D. Carlitz,
\prd{5}{1972}{3231.}  
E. Alvarez, \prd{31}{1985}{418;} \npb{269}{1986}{596.}
M. Bowick and L.C.R. Wijewardhana, \prl{54}{1985}{2485.} B. Sundborg,
\npb{254}{1985}{883.} S. N. Tye, \plb{158}{1985}{388.} E. Alvarez and
M.A.R. Osorio, \prd{36}{1987}{1175.}}

\lref\rcapa{N. Cabibbo and G. Parisi, \plb{59}{1975}{67.}} 

\lref\rrw{P. Salomonson and B. Skagerstam,
\npb{268}{1986}{349.} D. Mitchell and N. Turok, \prl{58}{1987}{1577.}
D.A. Lowe and L. Thorlacius, \prd{51}{1995}{665} \bb{9408134.}}

\lref\rkogan{I.I. Kogan, {\it JETP. Lett.} {\bf 45} (1987) 709.
 B. Sathiapalan, \prd{35}{1987}
{3277.}}

\lref\rgreeks{I. Antoniadis and C. Kounnas, \plb{261}{1991}{369.}
I. Antoniadis, J.P. Derendinger and C. Kounnas, \npb{551}{1999}{41,}
\bb{9902032.} I. Bakas, A. Bilal, J.P. Derendinger and K. Sfetsos,
\npb{593}{2001}{31,}
\bb{0006222.}}

\lref\rcorrh{M. Bowick, L. Smolin and L.C.R. Wijewardhana, {\it Gen. Rel.
Grav.} {\bf 19} (1987) 113. G. Veneziano, {\it Europhys. Lett.}
 {\bf 2} (1986) 199.
L. Susskind, \bb{9309145.} G. Veneziano, in {\it Hot Hadronic Matter:
Theory and Experiments}, Divonne, June 1994, eds J. Letessier, H. Gutbrod
and J. Rafelsky, NATO-ASI Series B: Physics, {\bf 346} (1995), p. 63.
A. Sen, \mpla{10}{1995}{2081.} E. Halyo, A. Rajaraman and L. Susskind,
\plb{382}{1997}{319,}
\bb{9605112.} E. Halyo, A. Rajaraman, B. Kol and L. Susskind,
\plb{401}{1997}{15,}
\bb{9609075.}}

\lref\rHP{G.T. Horowitz and J. Polchinski, \prd{55}{1997}{6189,}
\bb{9612146.}}

\lref\rusabel {S.A. Abel, J.L.F. Barb\'on, I.I. Kogan and
 E. Rabinovici,
\jhep{9904}{1999}{015,}
\bb{9902058.}}

\lref\rthresholds{J.L.F. Barb\'on, I.I. Kogan and E. Rabinovici,
\npb{544}{1999}{104,} \bb{9809033.}}

\lref\rmicro{R. Brandenberger and C. Vafa, \npb{316}{1989}{391.}
N. Deo, S. Jain and C.-I. Tan, \plb{220}{1989}{125;} \prd{40}{1989}{2646.}
N. Deo, S. Jain, O. Narayan and C.-I. Tan, \prd{45}{1992}{3641.}
}

\lref\rhpvd{G.T. Horowitz and J. Polchinski, \prd{57}{1998}{2557.}
T. Damour and  G. Veneziano,
\npb{568}{2000}{93,}
\bb{9907030.}}

\lref\rHPage{S.W. Hawking and D. Page, \cmp{87}{1983}{577.}}

\lref\rthorn{C. Thorn, \plb{}{1981}{458.}}

\lref\rghen{G.W. Gibbons and S.W. Hawking, \prd{15}{1977}{2752.}}

\lref\rgpy{D.J. Gross, M.J. Perry and L.G. Yaffe, \prd{25}{1982}{330.}}

\lref\rwithp{E. Witten, {\it Adv. Theor. Math. Phys.} {\bf 2} (1998)
505 \bb{9803131.}}

\lref\rmalda{J. Maldacena, {\it Adv. Theor. Math. Phys.} {\bf 2} (1998)
231 \bb{9711200.}}

\lref\rento{E. Alvarez and  T. Ort\'{\i}n,  
{\it Mod. Phys. Lett.} {\bf A7} (1992) 2889. }  

\lref\rncos{S.S. Gubser, S. Gukov, I.R. Klebanov, M. Rangamani and  E. Witten, 
{\it J. Math. Phys.} {\bf 42} (2001) 2749 
{\bb 0009140.} J.L.F. Barb\'on and E. Rabinovici, \jhep{0106}{2001}{029}  
\bb{0104169.}}   


\baselineskip=15pt

\line{\hfill CERN-TH/2001-375}
\line{\hfill RI-01/12-05}
\line{\hfill {\tt hep-th/0112173}}

\vskip 1.0cm

\Title{\vbox{\baselineskip 12pt\hbox{}
 }}
{\vbox {\centerline{Closed-String Tachyons and   }
\vskip10pt
\centerline{the Hagedorn Transition in AdS Space}
}}

\vskip0.3cm

\centerline{$\quad$ {\caps 
J.L.F. Barb\'on~$^{a,}$\foot{ On leave
from Departamento de F\'{\i}sica de Part\'{\i}culas da 
Universidade de Santiago de Compostela, Spain.} and 
E. Rabinovici~$^{b}$ 
}}
\vskip0.5cm

\centerline{{\sl $^a$ Theory Division, CERN, 
 CH-1211 Geneva 23, Switzerland}}
\centerline{{\tt 
barbon@cern.ch}} 

\vskip0.2cm

\centerline{{\sl $^b$ Racah Institute of Physics, The Hebrew University,  
 Jerusalem 91904, Israel}}
\centerline{{\tt eliezer@vms.huji.ac.il}}

\vskip0.2cm

\centerline{\bf ABSTRACT}

 \vskip 0.1cm

 \noindent 
We discuss some aspects of the behaviour of a string gas
at the Hagedorn temperature from a Euclidean point of view. Using
AdS space as an infrared regulator, the Hagedorn tachyon can be
effectively  quasi-localized and its dynamics controled by
a finite energetic balance. We propose that the off-shell RG flow
matches to an Euclidean AdS black hole geometry in a generalization of the
string/black-hole correspondence principle. The final stage of the
RG flow  can be interpreted  semiclassically as
the growth of a cool black hole in a hotter radiation bath.
The end-point of the condensation is the large Euclidean AdS black hole,
and  the part of spacetime behind the horizon has been removed.   In the
flat-space limit, holography is manifest by the system creating its own
transverse screen at infinity. This leads to  an argument, based on the
energetics of the system, explaining  why the non-supersymmetric type 0A
string theory decays into the supersymmetric type IIB vacuum.
We also suggest a notion of `boundary entropy', the value of which
decreases along the line of flow.

\vskip 0.5cm

\Date{December 2001}
               

\vfill





\baselineskip=14pt

\newsec{Introduction}

\noindent

The problem of the behaviour of strings at high temperature is a classic,
this on account
of its physical interest and its elusive aspects. The possibility that a 
Hagedorn spectrum \refs\rhag\ 
and its apparent limiting temperature are indications
that string theory has a new phase at high temperatures has been
extensively studied \refs{\rhist, \rkogan, \raw, \rrw, \rmicro, \rusabel}.

 In the Euclidean version the Hagedorn temperature can
be associated with the appearence of a thermal tachyon. The world-sheet
manifestation of closed-string target-space tachyons is the appearence of
relevent
operators which drive the system to a new infrared fixed point. In target
space such a flow will have an important impact on the space. In
particular it may change its dimension as the value of the Virasoro
central charge decreases. The infrared fixed-point string theory
should thus be described by some non-critical string background.    
Open-string tachyons have less of an impact; they do drive the system to a
new fixed point, however in that case the infrared fixed-point system has
the
same value of the
central charge as the ultraviolet fixed-point system.  The whole system
does loose
degrees of freedom along the flow; these are contained in the boundary
entropy function which does indeed decrease \refs{\rben,
 \rraben, \rrelevance, \rshatas, \rmarinho}.  

 Recently an intermediate type
of infrared instability was studied \refs{\raps, \rvafa,  \rchicago}.
 It contains a tachyon condensate
somewhat localized in a non-compact space-time. The central charge does
not change along the flow but the number of degrees of freedom does change
and a decreasing quantity associated with the flow is  identified in
orbifold examples \refs\rchicago.
 
One can be content with analyzing the end-points of the
flow. One may also wish to obtain a time-dependent picture of the
dynamics of the flow. In this note we set the  Hagedorn transition in
a time-dependent framework and offer a very qualitative picture of 
how the Hagedorn transition can be viewed as a flow of a `localized' tachyon.
To make this possible we will embbed the Hagedorn tachyon in an Anti de-Sitter
(AdS)  
background which serves as a regulator. It is a non-compact space in which
the temperature is redshifted as one travels along the radial direction.
The redshift localizes the amount of space for which the temperatures are
large enough to produce a thermal tachyon. We will describe several
competing  trajectories for such a flow, all of them will have a common 
end-point, a large stable black hole in AdS. In the process of reaching the fixed
point the Hagedorn temperature region is cut out of space, i.e.  the 
Euclidean black hole
is a region of space capped at its horizon. 
The central charge is retained in the flow as the dimension of space
remains unchanged. The extent of the dimension is diminished and the
amount of space gobbeled-up is related to the entropy change.

In section 1 we review and elaborate on  some facts associated with the
Hagedorn behaviour of strings. In particular we recall the string/black-hole
correspondence \refs{\rcorrh, \rHP, \rhpvd},
 which is vindicated in this note from an  Euclidean point of
view. In section 2 we recall mechanisms for black-hole nucleation
and discuss how the AdS space can serve as regulator which
will allow to expose a string/black-hole transition. In section 3 we recall
the emergence and the properties of thermal tachyons. In section 4 we
analyze possible
trajectories along which black holes will nucleate in hot AdS and evolve
to a stable black-hole configuration. In section 5 we give an argument
based on energy considerations why the non-supersymmetric type 0A string
background will decay into the supersymmetirc type IIB vacuum. It will
turn out that the Euclidean version of this decay is nothing but the above
decay of a hot AdS background into an AdS black hole. In section 6 we
suggest a geometric
version for the portion of the entropy that changes along the flow.

\newsec{General Aspects of String Thermodynamics}

\noindent

 The basic
observation about string thermodynamics 
is the fact that string excitations have
an asymptotic degeneracy of mass  levels that increases exponentially
as  
$\exp (\beta_s M)
$
for $\beta_s M \gg 1$,
where $\beta_s \sim \ell_s = m_s^{-1}$ is of the order of the string length.   
The 
Hagedorn temperature is defined as $ T_s =\beta_s^{-1}$
 and is naively interpreted as a limiting temperature for the string 
thermal ensemble. 

In QCD the string spectrum mimics that of hadronic resonances, and the
Hagedorn temperature $T_s$ is associated with the deconfinement temperature.
At $T>T_s$ the dynamics is better approximated by the quark-gluon constituents
\refs\rcapa.
In the context of 't Hooft's $1/N$ expansion, one can formalize these ideas
by saying that the thermodynamic entropy scales as $S=\CO(1)$ for $T<T_s$,
and $S=\CO(N^2)$ for $T>T_s$ \refs\rthorn.
  The possibility that a phase at $T>T_s$
may uncover fundamental degrees of freedom of critical string theory has
been the main physical motivation for looking at the Hagedorn behaviour
of `fundamental' strings. 

There are essentially two ways of looking at this problem. We could
roughly term them as `Hamiltonian' and `Euclidean'. In the first, one
works in the microcanonical ensemble, at fixed total energy,   and
 assumes that the
statistical mechanics of free strings is a good starting point, the
interactions being introduced adiabatically. In the second one focuses
on the formal evaluation of the canonical partition function at fixed
temperature $T$, i.e. the Euclidean path integral on $\CM
 \times {\bf S}_{\beta}^1$
with a thermal spin structure and $\beta = 1/T$.
 Here $\CM$ is the `spatial' conformal field
theory background,
that should contain a factor of ${\bf R}^{d-1}$ if the strings propagate
in $d$ large spacetime dimensions (we suppose that the remaining $10-d$
dimensions are of $\CO(\ell_s)$ length.) 

 In either formalism one has to face the
general problem of thermal ensembles in the presence of gravity, i.e.
the gravitational back-reaction of the finite energy density on
the background geometry. This occurs on scales of the Jeans length. For
a given total energy, the Jeans length corresponds to the radius of
the equivalent Schwarzschild black hole:
\eqn\jle{
R_J (E) = (G_d\,E)^{1 \over d-3},}
where $G_d $ is the $d$-dimensional Newton constant
$G_d \sim g_s^2 \,\ell_s^8 \,R_c^{d-10} \sim g_s^2 \,\ell_s^{d-2}$.
Alternatively, at fixed temperature $T$ the $g_{tt}$ component
of the metric develops a one-loop tachyonic  mass $m_J^2 \sim  -G_d T^{d}$ (c.f.
for example \refs\rgpy,) so that
\eqn\jlt{
R_J (T) = {1\over \sqrt{G_d\, T^d}}.}
In either case, we have to provide a physical `container' with volume
$V = R^{d-1} < R_J^{d-1}$ in order to maintain the framework of background  
flat geometry. For example, we may consider toroidal
compactification of the large dimensions to a torus ${\bf T}^{d-1}$ of
radius $R\gg \ell_s$. 
  Beyond this point, the physics depends on the particular
definition of the `container'.

   A good model for the
Hagedorn spectrum  is to imagine the highly excited strings as a random
walk of length $\ell_s^2\, M$ \refs\rrw.
One then finds that, even at the level of
free strings,  the  gas is
very sensitive to the finite-size effects introduced by the finite-volume
container \refs{\rmicro, \rusabel}.
 In general there is no notion of extensivity of the multi-string
thermodynamic quantities. The microcanonical density of states has the
form 
\eqn\microc{
\Omega(E)_{\rm free} = \exp (\beta_s E + \dots),}
where the dots stand for subleading contributions in the large-$E$ limit.
They can come in the form of lower powers of the energy balanced with
volume factors or in terms of logarithms. Since the purely linear term
in the entropy $S(E) = \beta_s E + \dots$ gives formally infinite
specific heat, the details of the thermodynamics depend on the subleading
terms. In any case, within this picture the Hagedorn temperature is
maximal within $\CO(1)$ accuracy in string units.  The microcanonical
temperature function
\eqn\mict{
T(E) = \left({\partial S \over \partial E}\right)^{-1}}
grows as $E^{1/d}$ for $T<T_s$, when the gas is dominated by massless
particles, and levels off at $T\approx T_s$ around $E_s
 = T_s (R T_s)^{d-1}$, where
$R$ is the linear extension of the finite volume of the system. 

Whether the function $T(E)$ is slightly decreasing or increasing for
larger energy densities depends on the subleading contributions to \microc.
These are model-dependent and can affect the physical definition of the
Hagedorn temperature by numerical factors of $\CO(1)$. On the larger picture,
we will see that these details are not very important. Therefore, again
  within $\CO(1)$ accuracy in string units, we can recognize
the regime dominated by long strings as a plateau of the microcanonical
temperature with $T\approx T_s$.

The interactions are introduced in this picture on the basis of some
general physical arguments. In principle, for a fixed total volume
$V = R^{d-1}$ there are two basic energy scales that depend on 
the string coupling
$g_s$. First, at energy densities  of order $1/g_s$,  
\eqn\plasma{E_{\rm plasma} =  {m_s \over g_s}\; (Rm_s)^{d-1} } 
 we can nucleate D-brane anti-D-brane pairs from the thermal
bath. Since D-branes are charaterized by
the fundamental open strings attached to them, it is natural to  assume that
the {\it leading} linear form of the entropy $S(E) \approx \beta_s E$ is
not affected. As for the subleading terms, the population of open-string
modes has the general effect of tilting the curve $T(E)$ slightly
upwards, i.e. open-strings work in favour of positive specific heat
\refs\rusabel.

In principle, we cannot exclude that such  relativistic plasma of D-branes 
 might be characterized by a different density of states, perhaps dominated
by the larger entropy of random $p$-branes with $p>1$ 
(see, for example \refs\rento.) However, we consider this unlikely on the
basis of the absence of a consistent quantization of relativistic $p$-branes
with $p>1$.

 The next important scale is
\eqn\gravs{
E_g = {m_s \over g_s^2}.}
At this energy the entropy in the string gas is of the same order of
magnitude as that of a Schwarzschild black hole:
\eqn\bhen{
S(E)_{\rm bh} = C_d\,E  \,(G_d\, E)^{1 \over d-3},}
where $C_d$ is a dimension-dependent numerical constant.
Although the details of the black-hole nucleation out of the
long-string gas are yet to be developed, we expect on general entropic
grounds that  the thermal ensemble is dominated by
a single large Schwarzschild black hole for $E>E_g$. In practice, this
is a generalization of the black-hole/string correspondence principle
\refs{\rcorrh, \rHP,
\rhpvd}. 
An idea of how this works can be obtained by a simple-minded estimate 
of the equilibrium  between a long-string `soup'  and a black hole
that nucleates out of a thermal fluctuation. The total entropy is of order
\eqn\est{
S(E,M)=\beta_s \,(E-M) + C_d\,M  \,(G_d\,M)^{1\over d-3},}
where $E$ is the total energy and $M$ is the mass in the black-hole phase.  
The function $S(E,M)$ decreases monotonically away from $M=0$ and has 
minimum at 
\eqn\eqcon{
M_c \sim {\ell_s^{d-3} \over G_d} \sim {m_s \over g_s^2} \sim E_g.}
Then, it increases monotonically for $M>M_c$. For sufficiently large 
total energy $E$, the global maximum is at $M=E$. Thus, if a black hole of
mass $M$ is nucleated by a thermal  fluctuation, it  will tend to
evaporate for $M<M_c$. On the other hand, if $M >M_c$,   it will tend to   
 grow without bound until it engulfs the
whole string gas with total available energy $E$. This process can be
very fast, as the small black holes should nucleate everywhere and
coalesce into a large black hole.  
Since the black holes have negative specific heat, the microcanonical temperature
decreases for $E>E_g$ and we find that $T\approx T_s$ is effectively a maximal
temperature of the system. Notice that this instability is physically
different from the Jeans instability, since it occurs at string-scale
distances. 

It should be emphasized here that the status of \eqcon\ as a {\it true} 
minimum of the entropy functional is not solid. Since the black hole with
mass $M_c$ has a Schwarzschild radius of the order of the string scale,
the entropy formula \bhen\ suffers $\CO(1)$ corrections  for $M<M_c$.
Thus, it is possible that there is no minimum at all and the long-string
gas is unstable at even lower `black hole' masses. On the other hand,
for $M>M_c$ the formula \bhen\ becomes more and more accurate and we
can trust the basic picture of the growing black hole.

The growth of the black hole continues until energies of order
\eqn\ehol{
E_{J} = {R^{d-3}   \over G_d}}
are reached, where the black hole crushes the walls of the container.
This is the holographic bound for this volume, $R=R_J$, and we cannot go further
in increasing the `energy' without jeopardizing the very notion of
energy by gross gravitational back reaction.

Although this picture is appealing from the point of view of the
holographic principle, there is no sign of phases with temperatures
substantially larger than $T_s$ and
no sign of `dissociation' of strings into constituents (a more conventional
picture arises for NCOS strings c.f. \refs\rncos.) Instead, black-hole
horizons shield all information about the structure of the string ensemble.
In particular, the transient Hagedorn plateau at $T \sim T_s$ even dissappears
for $E_s < E_g$, or
\eqn\limitg{
g_s^2 > (RT_s)^{1-d}.}
Therefore, the issue of whether string constituents are uncovered is
postponed until the physics of the finite-volume container is sorted out.
This is precisely what the AdS/CFT correspondence gives us.

\newsec{AdS Regularization}

\noindent

In this section we show that a physical specification of the `container'
by means of AdS space, together with the holographic interpretation of
such backgrounds,  
 yields a global physical picture that
realizes the analogy with QCD.

As a physical finite-volume regularization we take the ${\rm AdS}_5 \times
{\bf S}^5$ background of radius $R= \ell_s (g_s N)^{1/4}$. Type IIB string
theory in this background with $N$ units of RR flux through the ${\bf S}^5$ is
conjectured to be dual to $\CN =4$ SYM theory on the four-dimensional
boundary \refs\rmalda.
 We choose a timelike killing vector so that the metric reads
\eqn\met{
ds^2 = -dt^2\,\left(1+ {r^2 \over R^2}\right) +{dr^2 \over 1+ r^2 / R^2}
+ r^2 \;d\Omega_3^2 + R^2 \;d\Omega_5^2.}
We consider the thermal ensemble defined by $t\rightarrow -it$ and
$t \equiv t + \beta$. The parameter $\beta$ is  the inverse temperature
at $r=0$. The local temperature  $T/\sqrt{g_{tt}}$ then falls as $TR/r$ at
large $r\gg R$. On the other hand, on distance scales $\ell\ll R$ the
metric is well approximated by the flat ten-dimensional metric. Therefore,
hot AdS works like a `box' or container of size $R$, the radius of
curvature.  

The flat-space limit is thus $N\rightarrow \infty$ with fixed $\ell_s$ and
$g_s$. Notice that 
this is a difficult limit on the SYM side of the AdS/CFT duality,
since it takes to infinity both $N$ and the 't Hooft coupling $\lambda = g_s N$.
The Jeans length for a string-scale temperature inside the
container is
\eqn\jli{
R_J (T = T_s) = {\ell_s \over g_s}.}
Therefore, we cannot take the flat space limit without bringing the
Jeans instability into the interior of the `box'. Let us suppose for
the time being that $R_J > R$, so that no Jeans instability is present
in the system. This forces us to tune the string coupling as $N\rightarrow
\infty$ in the window:
\eqn\tune{
{1\over N} \ll g_s < {1\over N^{1/5}}.}

The main novelty of the infrared regulator is a change in the character
of the black holes for $r_0 > R$, where $r_0$ is the Schwarzschild
radius. In this regime,  black holes engulf the `box' and have positive
specific heat, with $r_0 \sim R^2 T$, energy $E \sim  N^2 \,(RT)^4 /R$ and
entropy $S\sim N^2 \,(RT)^3$ \refs\rHPage.
  They mimic the high-temperature
phase of a four-dimensional gauge theory. Thus, the large AdS black holes
represent, via the holographic correspondence, the deconfined phase. The
`constituent' degrees of freedom being the Yang--Mills gluons of the
dual theory.

For this system, the equivalent Jeans energy for the box is
\eqn\boxen{
E_J (R) = {R^{7} \over G_{(10)}} \sim {N^2 \over R}.}
However, larger energies are now  possible, and they  precisely correspond
to the large AdS black holes. The curve $T(E)$ can now be
 continued beyond $E_J$, and it rises as $E^{1/4}$, 
 exactly like a system with
positive specific heat in four spacetime dimensions.
 Notice that a Maxwell construction suggests that a canonical 
ensemble analysis will jump from the massless graviton phase directly to
the large AdS black hole via a first-order phase transition, skipping all
the transient phases
 dominated by Hagedorn strings and ten-dimensional black holes \refs{\rmarti,
\rabelr}.   
We have summarized the situation in Fig. 1. 

\fig{\sl The microcanonical temperature as a function of energy.
The Maxwell construction gives the critical temperature for the
Hawking--Page phase transition at $T_c \sim 1/R$.
}{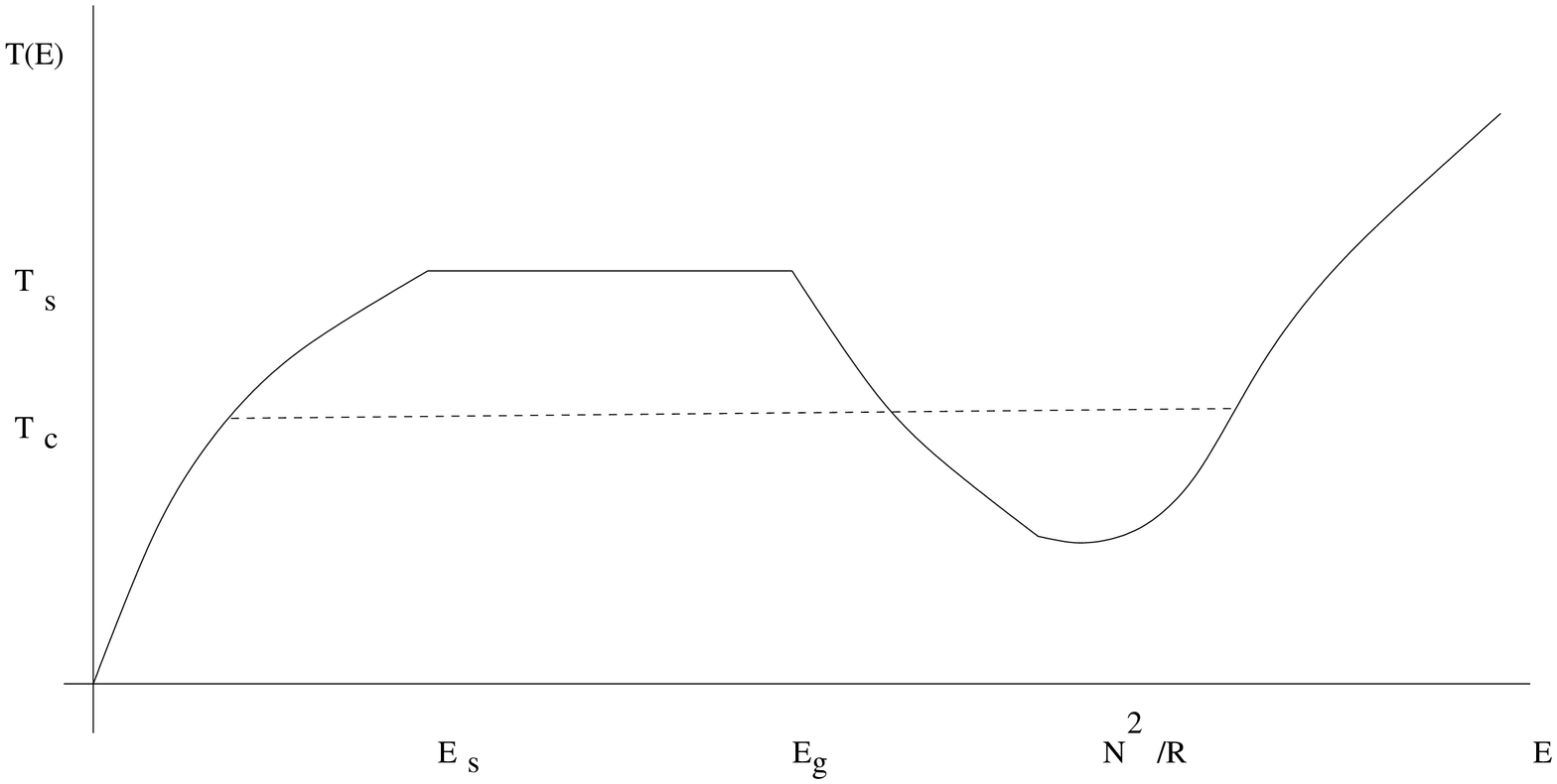}{5truein}

This agrees with  the canonical analysis of Refs. \refs\rHPage, \refs\rwithp. 
The AdS black hole dominates the thermodynamics
already for $T>T_c  \sim 1/R$. Since $R T_s \ll 1$ in the flat-space limit,
 the regime of Hagedorn temperatures
cannot be studied in terms of stable canonical configurations. In \refs{
\rusex, \rthresholds}\ 
 we termed this fact as a  `Hagedorn censorship'.
This means that long before  the Hagedorn temperature is reached
the large AdS black hole configuration already  dictates the asymptotic
equilibrium physics. The mechanism of the
transition may depend of the fate of metastable and non-stable
configurations. In this note we will present a qualitative picture of this
mechanism.

Even if the large AdS black hole dominates the thermodynamics above
the critical temperature $T_c \sim 1/R$, it is interesting to estimate  
the lifetime of the AdS `box' for $T>T_c$.  
 For temperatures 
just above the critical temperature  $T_c \sim 1/R \ll T_s$,
the relevant physics should be understandable in low-energy terms. There
are two low-energy instabilities of a massless gas in `approximately' flat
space. One is the Jeans instability and the other is the Gross--Perry--Yaffe
decay by microscopic black-hole nucleation (GPY) \refs\rgpy.
 The Jeans instability
is a one-loop effect, whereas the GPY instability is a non-perturbative effect.
Therefore, the Jeans instability is by far the dominant mode of decay whenever
present. Let us adjust the value of the string coupling as in \tune\
  so that, while having
a large AdS `box', we still have $R_J > R$ and the Jeans instability is
absent. Then the dominant decay mode might be
 the `sphaleron' process of GPY. These
authors interpret 
the Euclidean section of the Schwarzschild metric as a kind of sphaleron
 solution
that gives the leading semiclassical
 rate for thermal nucleation of black holes. 
  The rate is
suppressed by the non-perturbative factor 
\eqn\rategpy{
\exp\,\left(-C / G_d\,T^{d-2}\right).}
The nucleation produces a black hole inside the box, 
in unstable equilibrium  with the gas at
temperature $T>R$. This black hole is unstable, either evaporating back
into the thermal bath, or   growing  
 until it reaches the size of the large AdS black hole. 

The suppression factor \rategpy\ may seem to be  
of order $\exp(-C/g_s^2)$ at temperatures of $\CO(T_s)$. However, 
the calculation leading to \rategpy\ is not reliable at those temperatures,
since the relevant saddle-point black-hole geometry has Schwarzschild
radius $r_0 =\CO(\ell_s)$. Therefore, corrections of $\CO(1/g_s^2)$ 
to the exponent in \rategpy\ are possible by tree-level stringy effects. 

As an illustration of the rapid variation of the free energy as a function
of the Scharzschild radius, consider  the free energy  of a ten-dimensional
Schwarzschild  black hole of radius $r_0$ at temperature $T\sim T_s$.
Neglecting numerical factors, it has the form:
\eqn\estim{
 \beta F = \beta M - S \sim {1\over G_{(10)}} (\beta \,r_0^7 - r_0^8)
\sim {1\over g_s^2} \left((m_s r_0)^7 - (m_s r_0)^8\right).}
The GPY saddle point lies at the local maximum of this function,
at $r_0 = 7\ell_s /8$. However, at $r_0 = \ell_s$ we have $\beta F=0$
and at $r_0 = 2\ell_s$ we have a large negative value $\beta F=-2^7 /g_s^2$.

Because of this large sensitivity to the matching point around $r_0 \sim \ell_s$,
 we do not know {\it a priori} if the nucleation of a
string-sized black hole is actually suppressed. In section 3 we shall argue
that such nucleation is actually unsuppressed and
appears  as a {\rm tree-level} instability of
the string background.

\newsec{Hagedorn Behaviour and Tachyon Condensation}

\noindent

In the Euclidean approach one works at a more formal level by studying
the path integral on $X_\beta = {\bf S}_\beta^1 \times {\bf T}^{10-d}\times
{\bf R}^{d-1}$. In
order to evade obvious problems with Jeans instabilities we can replace
${\bf R}^{d-1}$ with ${\bf T}^{d-1}$ of radius $R$, with $\ell_s \ll R \ll
R_J$. 
The Hagedorn instability is seen here as the emergence of a tachyonic
winding mode around ${\bf S}_\beta^1$ precisely at $\beta < \beta_s $ 
\refs{\rkogan,\raw}. The
singularities of the thermodynamic functions are dominated by the
infrared singularities induced by the tachyon in string diagrams.  

In this picture, there is no formal objection to taking $T>T_s$, except
for the existence of the tachyonic mode. Therefore, if one could solve
for the dynamics of the tachyon condensation, one would have a
picture of the $T>T_s$ phase. In a mean-field Landau--Ginzburg 
approach one would seek a static effective action
for the tachyon field $\sigma$ that gives the following contribution
to the Euclidean action or free energy: 
\eqn\tacp{
I_{\sigma} =
 {\beta_s \over 2G_d} \int d^{d-1} x \,\left[ f(|\sigma|)\,
|\partial \,\sigma|^2
+  m_s^3 \,(\beta-\beta_s) \,|\sigma|^2 + m_s^2\, V(|\sigma|) + \dots\right].}
Notice that both positive and negative winding modes contribute, making the
Hagedorn tachyon into a
 complex scalar field. The dots stand for higher derivative terms generated
at tree level or other contributions generated at higher loops or even
non-perturbatively (c.f. for example \refs\rbvm.) 
 
It was argued in \refs\raw\
that the interaction with the dilaton  affects $V(|\sigma|)$ so that
one has actually a first-order phase transition at a slightly smaller
temperature than $T_s$. In any case, 
 since $V(|\sigma|)$ is already non-trivial at the tree
level, the dynamics of the tachyon destabilizes the system very fast,
and the quasi-static approximation is not appropriate.
 In particular, free energies of $\CO(1/g_s^2)$ are released in the
process of condensation.

A standard technical assumption is that   the  real-time dynamics of  
 the tachyon condensation
process might be mimicked by some  
 renormalization-group  (RG) flow on the world-sheet. 
 Thus,  we speak of RG flow metaphorically, 
 as if it was
real time evolution, but we just have a collection of off-shell
string backgrounds. For example, the failure to satisfy the
tachyon equation of motion is just the beta function for the
corresponding world-sheet coupling:
\eqn\betaf{
\beta_\sigma \propto {\delta I \over \delta \sigma}.}  

 One picture for the  required RG flow  
 was developed in the
early days in terms a Kosterlitz--Thouless transition on the world-sheet
 \refs\rkogan.
In this picture the tachyon vertex operator
 is a world-sheet vortex and its condensation
renders   the Euclidean-time world-sheet field massive. Since one dimension
`dissapears' it  appears that the end-point of condensation must be  
a non-critical string with a Liouville field taking the role of `time', as
in 
\rliouville. Although
this seems like the   natural set-up for the real-time description of the
condensation  process,
the lack of a clear
 thermodynamical interpretation of the end-point is rather puzzling.  

At any rate, the end-point of the condensation
of the Hagedorn  tachyon is guaranteed to have a non-trivial physical
interpretation, if only because the tachyon itself is not a physical
state of the string theory at zero temperature. Thus, it is not clear
even in principle what the expectation value of the $\sigma$-field means.  
This is somewhat reminiscent of similar situations in the study of
the decay of 
non-BPS branes to a pure closed-string vacuum \refs\rsen:
 one needs some dynamical
mechanism, presumably operating at the tree level, to suppress unwanted
excitations at the condensed vacuum. In the case of the Hagedorn tachyon,
such a requirement could be incorporated by proving that the kinetic
function $f(|\sigma|)$ in
\tacp\ has a zero at the end-point of tachyon condensation.

One striking result is that of Refs. \refs\rgreeks. In these works the
tachyon potential, together with other moduli, is argued to be calculable
on the basis of supersymmetry constraints. Since one finds (Euclidean)
 vacua with
restored supersymmetry, it is natural to view them as end-points of
tachyon condensation in close  analogy with the behaviour of
localized tachyons on D-branes or orbifolds \refs{\rsen,\raps}.
  This view is also vindicated by recent conjectures
on the fate of the closely related Type 0 string backgrounds \refs\rtipocero.
 These would
decay into supersymmetric type II string backgrounds \refs{\rcosta, 
\rgutstr, \rgutperlem}.

In any case, a decay of the canonical ensemble geometry into a background
with supersymmetry (typically flat space or some domain wall solutions) is
also difficult to interpret in terms of the original statistical mechanical
problem, especially  in view of the picture outlined before
in  the microcanonical ensemble.

In view of the previous remarks, 
perhaps the most striking fact about the canonical approach is the lack
of a clear physical picture. In fact, it offers no obvious candidate
for realizing the `QCD scenario', i.e. the liberation of `string 
constituents'.  In this paper we make an  attempt at  providing such a 
physical picture.

\newsec{Euclidean Black Holes and the  Matching Hypothesis}

\noindent

The idea that closed-string tachyons destabilize the vacuum and drive it
into a new vacuum has appeared in the past in various settings. In the
simplest form a spacetime coordinate becomes massive on the world-sheet
 and disappears
in the deep infrared leading to an effective reduction of the dimension
of spacetime \refs\rkogan, or crystalization and formation of asymptotic
curvature \refs\refr. 
 While this flow has a certain appeal its main drawback is
that it forces us into the setting of a non-critical string theory
 in which little progress has been made so far.
However one may consider a situation where the effect on space of the flow
is important but less drastic. The dimensionality of space does not change
in the infrared but nevertheless the extent of this dimension is reduced.
This can  happen in a setup of localized tachyons whose flow does not
change the bulk central charge but does does change some parameter counting
a lower dimensional change in the degrees of freedom \refs{\raps, \rchicago}.

In this section we argue that  such ideas find a natural implementation
in the AdS regularization of the Hagedorn behaviour. 
Let us consider the  ${\rm AdS}_5 \times {\bf S}^5$ thermal  background
at temperature slightly above the Hagedorn limit
$\beta < \beta_s$. For $r < R$ we can consider the thermal
manifold as well approximated by ${\bf S}_\beta^1 \times {\bf R}^9$.
On the other hand, for $r>R$ the temperature drops below the string scale,
 so that thermal
winding modes are never tachyonic in the asymptotic region. 
Therefore, we can use the AdS/CFT background as an infrared regulator
that regards the Hagedorn tachyon as a `quasilocalized' tachyon.

For this quasilocalized tachyon there is an obvious generalization of
the picture developed in \refs\raps.
 Since spacetime for $r\ll R$ is
an approximate cylinder with a slightly $r$-dependent opening angle,
it is natural to suppose that it behaves similarly to a narrow cone
such as those obtained from a ${\bf C}/{\bf Z}_n$ orbifold for 
large $n$. In that case, there would be an effective capping of the
cone propagating towards large $r$.  In our case
 we are actually looking for static
Euclidean interpolating backgrounds that  decrease the free energy.
Hence, it is natural to assume that the appropriate configurations are a set
of  backgrounds labelled by $r_c >0$, with an effective string-scale
capping around radial coordinate $r = r_c$.  In principle, the free
energy of such backgrounds should be defined from the world-sheet 
CFT description,
 or perhaps in the holographic dual SYM theory.  

A more formal statement would be that there is a field redefinition
with the property that the  tachyon expectation value
is nothing but the value of the radial variable $r_c$ at which the geometry
is capped:
\eqn\expt{
|\sigma|  \sim m_s \,r_c.} 
The decay of the tachyonic part of the manifold proceeds primarily
 via a tree-level
RG flow on the world-sheet. For example, one interesting concrete
mechanism is some analogue of 
 the Kosterlitz--Thouless flow of Refs. \refs\rkogan. For
inverse temperature $\beta < \beta_s$, the $r$-dependent size of
the thermal circle ${\bf S}^1$ is
\eqn\sizec{
\beta(r) = \beta\,\sqrt{1+ r^2/R^2}.}
As a rough 
estimate of the capping coordinate, we may define it by $\beta(r_c) =
\beta_s$. In the flat-space limit $R\gg \beta_s$ this  yields
\eqn\cappp{
r_c (\beta) = R\,\sqrt{\left({\beta_s \over \beta}\right)^2 -1}.}   
This formula shows that the capping distance can be kept smaller than
$\CO(R)$ only in a scaling limit.  
 Namely, unless $\beta_s -\beta$ is fine-tuned to be very small, the
stringy  capping
occurs around $r_c = \CO(R)$, i.e. the radius of the AdS/CFT `box'.
In order to estimate the free energy released in this process we need
a concrete geometrical model  for the spacetime decay.   

The Euclidean section of a black-hole spacetime has an effective
capping of the geometry at the radial coordinate of the horizon. From
the discussion in section 1, we know  
that the process of black-hole nucleation and growth is expected to
play an important  role in the exit from the Hagedorn regime.  Given
these two facts,  it seems
very natural 
 to regard the Euclidean black-hole geometry 
 as an effective geometrical model for the capping of the thermal cylinder.

Therefore, our main  hypothesis is simply that the stringy 
geometry resulting from the off-shell
tachyon dynamics with $\sigma \sim m_s r_c$ matches  onto an Euclidean  
AdS black-hole  geometry  with  Schwarzschild radius
\eqn\march{
r_0 \sim r_c (\beta).}

The black-hole geometry with horizon at $r=r_0$ is:
\eqn\sbm{
ds^2 = \mu(r_0, r)\,dt^2  +
{dr^2 \over \mu(r_0, r)} +  r^2 \,d\Omega_3^2 + R^2\,
d\Omega_5^2,}
where
\eqn\mufunc{
\mu(r_0, r) = 1+{r^2 \over R^2} - \left(1+{r_0^2 \over R^2}\right) \,{r_0^2
\over r^2}.}
Our choice of an AdS black hole rather than a ten-dimensional 
Schwarzschild black hole is motivated by our desire of maintaining  
the $SO(6)$ symmetry  that is inherited from
the ${\bf S}^5$ factor. This is a symmetry of the initial conditions,
since we were assuming uniform thermal
ensemble on ${\rm AdS}_5 \times {\bf S}^5$. The dynamical
capping  
of the geometry may proceed by initial nucleation  ten-dimensional
 black holes  that `coalesce' into a uniform AdS black hole with
$SO(6)$-symmetric horizon. As a technical simplification, we consider
only the $SO(6)$-symmetric manifolds \sbm\ for which we have
an explicit analytic form. 

  Notice that, according to the
remarks in section 2,   the GPY `sphaleron barrier' \rategpy\ suffers
from an $\CO(1)$ ambiguity at string-scale temperatures. In fact, our
mechanism suggests that at $T \geq T_s$ the barrier is
 completely washed out by the
tree-level dynamics of the Hagedorn tachyon \tacp.  On the other hand,
at temperatures not too far {\it below} $T_s$, it is tempting to
say that  the Atick--Witten 
 first-order phase transition described in \refs\raw\  and the GPY
nucleation are one and the same physical process.   
 
We now take \sbm\ as a model for the capped geometry and investigate
the Euclidean free energy of these manifolds.

\subsec{Off-Shell Black Holes}

\noindent

The free energy of the AdS black hole reads 
(in the following we rely heavily on formulas from \refs\rwithp):
\eqn\educ{
 \beta \,F = \beta \,M - S,}
where $M$ is the  mass of the black hole:
\eqn\adsms{
M = {3 {\rm Vol} ({\bf S}^3) \over 16\pi G} \left({r_0^4 \over R^2} + r_0^2 
\right),}
normalized to the AdS vacuum manifold, and $S$ is the Bekenstein--Hawking
entropy:
\eqn\ent{
S= {{\rm Vol}({\rm Horizon}) \over 4G}
 = {{\rm Vol}({\bf S}^3) \over 4G} \;r_0^3.}
The free energy can be computed as the difference of Euclidean actions 
between the black hole manifold $X_{r_0}$ and the thermal AdS manifold
$X_0$:
\eqn\dif{
\beta F = \Delta I(r_0) = I\left[X_{r_0} \right] - I\left[X_0\right].}
 The Euclidean action has the form (c.f. \refs\rghen): 
\eqn\eucl{
I(X) = -{1\over 16\pi G} \int \left(\CR +{6 \over R^2}\right)
 - {1\over 8\pi G} \int_{
\partial X} \CK,}
where we use the effective five-dimensional gravity action, with
$G \sim g_s^2 \ell_s^8 / R^5$. These formulas should provide a good
quantitative estimate of the free energy for black holes that
are sufficiently large in string units: $r_0 \gg \ell_s$.

 Notice that the black-hole capping changes the  topology of the 
space-time manifold. While
$X_0$ is homeomorphic to ${\bf S}^1 \times {\bf R}^4 \times {\bf S}^5$,
the topology of $X_{r_0}$ is ${\bf R}^2 \times {\bf S}^3 \times {\bf S}^5$.  
 In fact, $X_{r_0}$ with metric  \sbm\ is not a solution
of the equations of motion in general, 
because it has   a conical singularity at
$r=r_0$ with deficit angle:
\eqn\eeef{
\delta = 2\pi \left(1-{\beta \over \beta (r_0)} \right),}
where $\beta(r_0)$ is the inverse Hawking temperature of a black hole with
radius $r_0$:
\eqn\iha{
\beta (r_0) = {2\pi R^2 r_0 \over 2 r_0^2 + R^2}.}
 The conical singularity contributes a finite amount to the Euclidean action of
the black hole:
\eqn\eucl{
\Delta I =  \Delta I_{\rm bulk}  + I_\delta.} 
 The contribution proportional to the deficit angle comes from a delta-function
singularity in the curvature scalar and is proportional to the themodynamic
entropy:
\eqn\curfs{
I_\delta =  -\delta
\;{S \over 2\pi}.} 
The bulk contribution is proportional to the volume difference between
the two manifolds
\eqn\bulk{
\Delta I_{\rm bulk} ={1\over 2\pi G}\;\left[
 {\rm Vol}\left(X_{r_0} \right) - {\rm Vol} 
\left(X_0\right)\right].}
 In evaluating \bulk, there is a  subtlety in that the
volume of the black-hole manifold $X_{r_0}$ must be  computed with a rescaled
metric that takes into account the slight difference in the Euclidean-time
 periods,
when compared at large cutoff radial scale $L$.
Thus, the volume of the black-hole manifold comes with an extra factor
of
\eqn\extras{
{\beta' \over \beta} = \sqrt{\mu(0,L) \over \mu (r_0, L)}=
 1+ {R^2 \over 2} \,\left(1+ {r_0^2 \over R^2} \right) \;{r_0^2 \over L^4}
+ \dots}
Although this is a negligible  contribution to the action {\it density}
in the limit where we remove the cutoff $L\rightarrow
\infty$, it does contribute to the volume integral a finite term. This
finite term is not very important for $r_0 \gg R$,  
 but it is of relative $\CO(1)$  for $r_0 < \CO(R)$.

For a given temperature
 $T=1/\beta$, larger than the critical temperature $T_c$,
 there are two stationary points at which
$\delta =0$. They correspond to the two 
  roots of the equation
\eqn\earr{
\beta = {2\pi R^2 r_0 \over 2r_0^2 + R^2}.}
For $\beta \ll R$,  we have a small AdS black hole with $r_0 =\CO(\beta)$
and a large AdS black hole with $r_0 =\CO(R^2 T)$. The small black hole
corresponds to a local maximum of $\Delta I$ and it is
unstable, with negative specific heat. The large AdS black hole
corresponds to the global minimum of $\Delta I$ and it 
is stable,  with positive specific heat.  

All off-shell black holes with $r_0 < \CO(\beta)$ and $r_0 > \CO(R^2 T)$ are 
hotter than the thermal radiation and thus evaporate back to lower radii. 
On the other hand, all off-shell 
 black holes with $\CO(\beta)  < r_0 <\CO(R^2 T)$ are
cooler than the radiation and tend to grow. The coolest black hole 
has $r_0 = R/\sqrt{2}$ and deficit angle
\eqn\defma{
(\delta)^{\rm max} = 2\pi -{\beta \over R}
\left(\sqrt{2} + {2 \over \sqrt{2}}\right).}

For large deficit angles 
 the geometry is that of a very narrow cone that starts at $r=r_0$
and has length of $\CO(r_0)$. At this point it merges into a smooth geometry
with curvature of $\CO(1/R^2)$. Note that the part
of the free-energy release that comes from the progressive `opening' of
the cone can be 
literally described in terms of the  mechanism in 
 \refs\raps. 
The evidence from \refs\rdabho\ is that such stringy decay  
of the cone respects the semiclassical form of the thermodynamic functions,
provided $r_0 \gg \ell_s$. 

We notice however that in this problem one cannot simply model the
black-hole manifold  as a flat cone. In fact, the balance between the
conical \curfs\  and bulk \bulk\ contributions  to the free energy is such that
for small off-shell black holes $r_0 <R$, the complete free energy
decreases while the deficit angle actually {\it increases}. It is only
for $r_0 >\CO(R)$ that both contributions to the free energy decrease in
the process of black-hole growth.    

\fig{\sl Schematic plot of the free energy of AdS black holes. The dashed
line is the free energy at temperatures slightly above the critical
temperature  $T_c = \CO(1/R)$. The local maximum is the small AdS
black hole, which is the  GPY saddle point for this space,
and  the global minimum is the large AdS black hole. The full line gives the
free energy at $T > T_s \gg 1/R$. The circle indicates the matching
point where string effects become dominant. It is assumed that these
stringy effects wash out the GPY saddle point and at $r_0 \sim 0$ one
sees the tree-level instability of the Hagedorn tachyon.
}{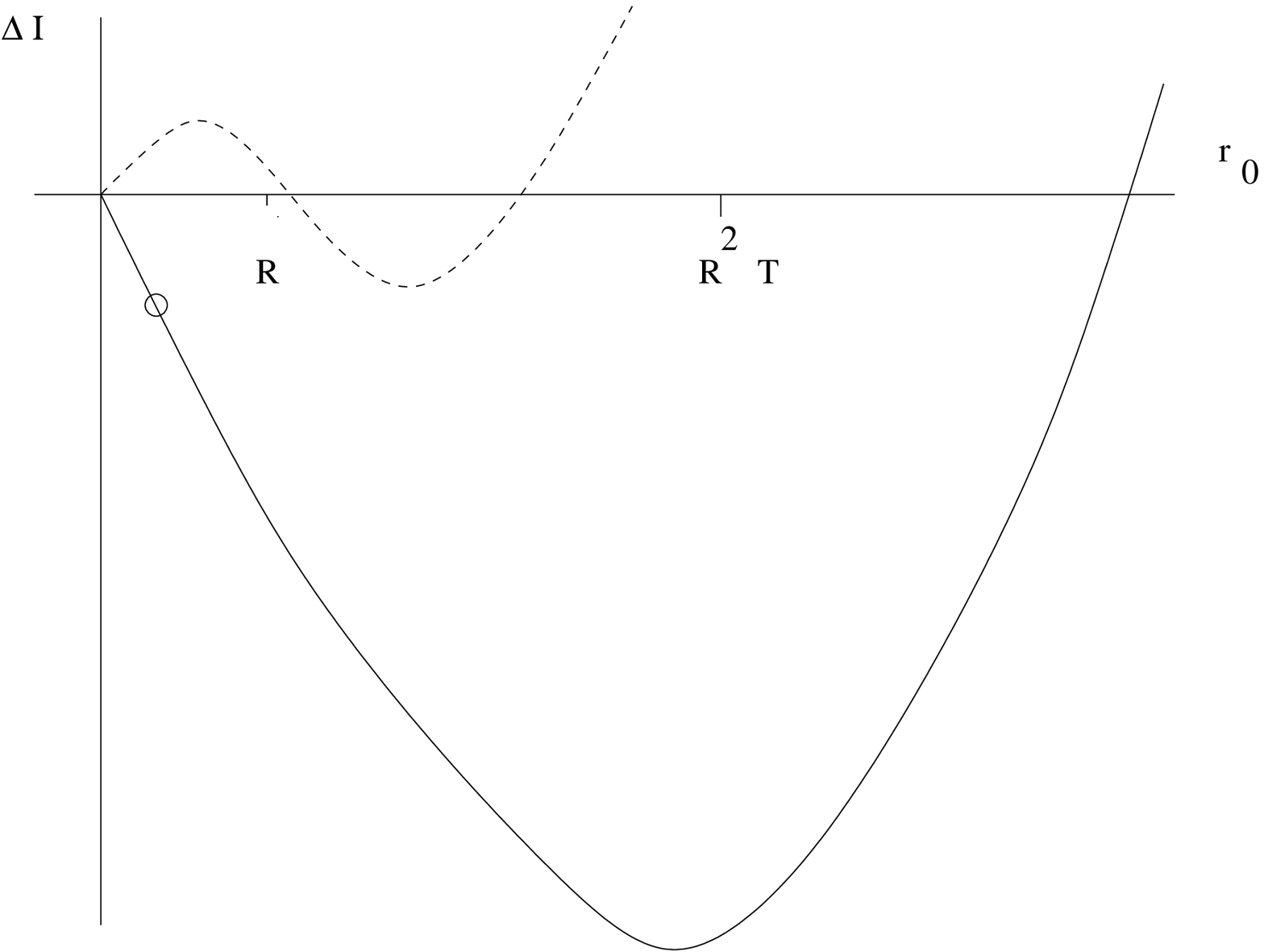}{5truein}

\subsec{The Semiclassical Regime of the Tachyon Roll} 

\noindent  

In order to trust the low-energy formulas for the free energy we must
have a matching point $r_0 \sim r_c (\beta) \gg \ell_s$. Since   
$\beta \sim \ell_s$ and $\ell_s \ll r_c (\beta) < \CO(R)$, we find
that the deficit angle at the matching is approximately maximal
\eqn\maxdef{
\delta_c \approx 2\pi.}
This means that the stringy process of decay matches at $r_c (\beta)$
to a very off-shell black hole that is very unstable 
towards growth, so that the RG flow continues semiclassically. 
The amount of free energy released at the moment of the matching is
of order:
\eqn\releasem{
\Delta I_{\rm matching} \sim -{r_c^3 \over G}.}
For $r_c = \CO(R)$, this yields
\eqn\ree{
\Delta I_{\rm matching} \sim -{R^8 \over g_s^2 \ell_s^8} \sim N^2.}

In the semiclassical stage of the flow, for $r_0 > r_c (\beta)$,  the
beta function of the tachyon is proportional to the deficit angle:
\eqn\proppp{
\beta_\sigma \propto {d I \over d r_0} = -{3 {\rm Vol}({\bf S}^3)
r_0^2 \over 8\pi G} \;\delta.} 
Therefore, the RG flow continues until the black hole grows into a
large, stable AdS black hole at 
 at $r_0 \sim R^2 T \sim \ell_s \,\sqrt{g_s N}$. At this point the deficit
angle vanishes and so does the beta function. 

Thus, our physical picture is the following. For $1\ll TR \ll T_s R$ the only
channel for decay into the large AdS black hole is the GPY nucleation,
a process that is exponentially suppressed (we are assuming that the
scales were chosen so that there is no Jeans instability.) Hence,
even if the large AdS black hole is the dominant thermodynamical
configuration, the vacuum AdS manifold is still long-lived for small
string coupling.

As the Hagedorn temperature is approached, the GPY barrier starts to
feel the string $\alpha'$ corrections. We conjecture that the free energy
as a function of the black-hole radius matches to the tachyon potential,
so that the first-order phase transition described in \refs\raw\ would be
simply the GPY nucleation when continued into the stringy domain. Finally,
for  $T \geq T_s$, the GPY barrier dissappears and
the  Hagedorn tachyon condenses via
a tree-level (and therefore fast) process.  
The decay can be followed semiclassically
for $r_0 >r_c (\beta)$ as the
 growth of a cool black hole in a hotter medium. Thus,
at the Hagedorn  temperature,
 the formation of the large AdS black hole becomes
a classical process and the thermal AdS manifold is short-lived.
We summarize the situation in Fig. 2.

\subsec{The End-Point of the Tachyon Condensation}

\noindent

The main prediction
of this scenario is that the expectation value of the tachyon at the true
minimum is
\eqn\expv{
\bra |\sigma| \ket \sim m_s \,r_0 \sim \sqrt{g_s N}.}
This is hierarchically larger than the value of the tachyon at the end of
the `stringy' condensation process, which was $\CO(m_s r_c) = 
\CO( (g_s N)^{1/4})$. Also, the total free energy released in the
complete process  is of order  
\eqn\totf{
\Delta I_{\rm total} \sim -{R^6 T^3 \over G} \sim -N^2 \,(RT)^3.
}

The possibility of having {\it bona fide} $\sigma$-field fluctuations around
the true vacuum can also be addressed here. Since the supergravity
interpretation of the $\sigma$ zero mode is the horizon location, the
frequency of oscillations of the stable end-point is proportional to
$d^2 I / d r_0^2 \sim R^2 /G$, when  
calculated around the stable minimum. 
This   diverges in the flat space limit $N\rightarrow
\infty$ at fixed $\ell_s$ and $g_s$. Thus, in the flat-space limit the
$\sigma$ fluctuations are frozen in the stable vacuum.

Finally, the apparent emergence of supersymmetric vacua as candidate
end-points of tachyon condensation can be given a simple thermodynamic
interpretation in our model. The basic observation is that the Euclidean
manifold corresponding to the final AdS black hole is simply connected.
Therefore, the spin structure is unique and at the horizon fermions
are locally smooth and in particular periodic in the natural polar
coordinate centered at $r=r_0$. This is a general feature of any Euclidean
black-hole metric in a supersymmetric theory, i.e. black holes
are always `supersymmetric' in the vicinity of the horizon capping.

Now, in the flat space limit $N\rightarrow \infty$ the curvature at
the horizon of the large AdS black hole goes to zero and we are
left with a better and better approximation of ${\bf R}^{10}$ with
the standard spin structure. Thus, in this limit we would interpret
the end-point of the decay process as supersymmetric (Euclidean) type IIB
spacetime, but sitting at the transverse boundary of the original spacetime.

This is a remarkable manifestation of holography which
could have important applications to other patterns of decay of
non-supersymmetric backgrounds.
In the next section we outline the first such application. 

\newsec{The Decay of 0A into IIB}

\noindent

The previous argument about the local supersymmetry of the large AdS black
hole around the horizon can be used to give an elementary proof of the
decay of a type 0A string theory into a supersymmetric type IIB model.
More exactly, we establish the decay of the corresponding Wick-rotated
theories in the sense that we have discussed in the paper.

The basic point is that the type 0A theory is formally defined as the
high-temperature limit of IIB theory, via standard T-duality 
\refs{\rrohm, \raw, \rgaberdiel}:  
\eqn\dual{
{\bf 0A} = \lim_{\beta \to 0} {\bf IIB}_\beta,}
where ${\bf IIB}_\beta$ stands for the type IIB theory on ${\bf S}^1_\beta 
\times {\bf R}^{9}$ with thermal boundary conditions.
But we just argued that the decay of  ${\bf IIB}_\beta$ can be regularized
as the decay of ${\rm AdS}_5 \times {\bf S}^5$ at high temperature. In fact,
even if our picture of the tree-level flow is not right, there is little
doubt that this background decays into the large AdS black hole, a
consequence of holography. But then, either in the flat-space limit
of AdS or in the large-temperature limit, the Euclidean black-hole manifold
cannot be locally distinguished from flat ${\bf R}^{10}$. 
The RR field-strength 
through ${\bf S}^5$  scales as $F_{\rm RR} \sim N/R^5$. Thus, it 
vanishes locally  as $N^{-1/4}$  in the flat-space
limit.  

Therefore, we have shown that indeed the Wick-rotated type 0A background
does decay into the type IIB supersymmetric background. If we now perform
a Wick rotation back to Lorentzian signature in some spectator dimension, 
(different from the compact supersymmetry-breaking dimension of the type 0A)
we have a similar statement about the real-time theories.

Other arguments in favour of this decay have appeared in the
 literature, \refs{\rgutstr, \rgutperlem}.
 We believe that the main advantage of our argument is  that it
provides a clear regularization that also gives an interpretation in
terms of energy balance. In some sense, such regularization is necessary
in order to be able to pose the question in precise terms.

\newsec{Boundary Central Charge and Local Central Charge}

\noindent

In \refs\rchicago\ a  notion of `boundary central charge' 
was introduced in the study of orbifold tachyon condensation. It is
supposed to play a role similar to that of the boundary entropy \refs{\rben,
 \rraben, \rrelevance, \rshatas, \rmarinho}\ 
 in open-string boundary RG flows. Namely, it should measure the decrease
in the effective number of degrees of freedom along the RG flow, in those
cases where the central charge does not change. 
The formal definition of the boundary central charge $g_{cl}$ is in terms
of the twisted torus partition function on the orbifold:
\eqn\tpart{
\lim_{\tau_2 \to 0} \;{\cal Z}(\tau, {\bar \tau})_{\rm twisted} \propto g_{cl}
\,\exp (\pi c /6\tau_2).}
After factoring out by the geometric volume of the orbifold fixed point,
the  quantity $g_{cl}$ is extracted:
\eqn\defg{
g_{cl} = {1\over |\Gamma|} \sum_{g\neq 1}{1\over N_g},}
where $N_g$ is the number of fixed points in the sector twisted by the
group element $g$ and
$\Gamma$ is the orbifold discrete group. The authors of \refs\rchicago\
have computed this quantity explicitly for various orbifold CFTs and
find agreement with the conjecture that $g_{cl}$ decreases along RG
flows. In particular, for ${\bf C}/{\bf Z}_n$ one gets
\eqn\exaa{
g_{cl} (n) = {1\over 12} \left(n-{1\over n}\right),} 
which indeed increases with $n$. Thus, such $g_{cl}$ would decrease
for the flows in \refs\raps\ that decrease $n$.  

We can give a physical interpretation
of \exaa\ as follows. From the original definition \tpart\ we see that
$g_{cl}$ has a factor of the available volume for propagation of
twisted strings in target space. Since we are factoring out the trivial
geometrical volume of the fixed-point submanifold, there remains the
effective volume given by the spread of the wave function of twisted
states in the directions transverse to the fixed-point submanifold.  

For the case of a cone, we can estimate the wave-function support of
the lowest-lying tachyonic state 
 by noticing that a   
twisted string is a winding mode around the tip of
the cone. The ground-state string  (non-oscillating)
  is forced to be localized near the tip because dragging it
out  costs at least the static energy of the wrapping around the base of the  
cone.  If we consider a cone of deficit angle $2\pi (1-1/n)$ 
 with base-length of $\CO(\ell_s)$, the
volume is $\pi \ell_s^2 \, n$. Thus, this argument captures
 the leading large-$n$
asymptotics of \exaa. 

Therefore, we propose a physical interpretation of $g_{cl}$ 
as a measure of `depleted volume' by the condensation process or, in
other words, a measure of the extension of the capping.

For the case of the thermal manifold ${\bf S}^1 \times {\bf R}^9$, the
analogue of the twisted partition function \tpart\ is the integrand
of the one-loop free energy in the sectors with non-trivial thermal
winding. Therefore, the analogue of $g_{cl}$ is given by effective
volume   available for winding modes, in this case all ${\bf R}^9$.
However, for the case of thermal  ${\rm AdS}_5 \times {\bf S}^5$, thermal 
winding modes are effectively confined to the `box' of radial extent
$r<\CO(R)$. For a capped manifold at $r=r_c$ we have, within $\CO(1)$
accuracy in string units:
\eqn\volc{
{\rm Vol} (X)_{r_c}^{R} = {1\over 4} \,\beta \,{\rm Vol}({\bf S}^3) \,
{\rm Vol}({\bf S}^5) \,R^5\,(R^4 - r_c^4).}
By analogy with the case of the cone, we may divide by the volume of
the `box' orthogonal to the $(t,r)$ space, which is of $\CO(R^8)$. Thus,
we obtain the following quantity in string units:  
\eqn\anal{
g_{cl} \sim (g_s N)^{1/4} \,\left(1-{r_c^4 \over R^4}\right).}
This expression is valid for $\ell_s \ll r_c < \CO(R)$. The overall scale
for such boundary central charge is set by the radius of curvature of
the AdS space in string units. It is clearly decreasing with the RG flow
if this is interpreted as the extension of the capping to larger radii. 
For $r_c > R$ the value of $g_{cl}$ should  
decay exponentially, since the thermal
winding modes become very massive in this region.  

The information about the RG flow contained in $g_{cl}
$ comes from a one-loop diagram in closed-string theory. In fact, it is
related to the one-loop correction to the free energy. 
The analogous quantity at tree level is thus the classical free energy that
we have studied in the previous section.  
In this sense, it is very interesting that at least the  bulk contribution  
to $\Delta I$ is essentially given by a measure of the `depleted volume':
\eqn\ass{
\Delta I = I_\delta + {1\over 2\pi G R^2}\,
\left[{\rm Vol} (X_{r_0}) - {\rm Vol} (X_0)\right],}
 Although $\Delta I$ is monotonically decreasing between the small and large
stationary black holes, for $\beta \ll r_0 \ll R^2 T$,
 neither the bulk piece nor the conical singularity
contribution are separately monotonic in this range. 

The bulk contribution is actually monotonically {\it increasing} for
$\CO(\beta) < r_0 < \CO(R)$, because  of the 
 offset volume effect at large radius of
Eq. \extras.  On the other hand, $I_\delta$ reaches its minimum at
$r_0 = \CO(R^2 T)$ but at a value strictly smaller than the Schwarzschild
radius of the large AdS black hole. This is easy to understand since
$I_\delta =0$ at the large stable AdS black hole.  

It might be interesting to consider  alternative 
 splittings of the Euclidean action. For example, instead
of distinguishing between the volume contribution and the conical
singularity, we can split the black-hole manifold  in two pieces by
cutting out a small cone $D_\epsilon$  of radius $\epsilon$ around the horizon
in $(r,t)$ space.  
Thus, $X= X_\epsilon + (X-X_\epsilon)$ with $X_\epsilon = 
D_\epsilon \times {\bf S}^3 \times {\bf S}^5$. Then, the contribution
to $\Delta I$ from $X_\epsilon$ (with due attention to the contribution
of the extrinsic curvature term to the action,) gives minus the 
thermodynamic entropy:
\eqn\comen{
-S = \lim_{\epsilon \to 0} \left[ -{1\over 16\pi G}
\int_{X_\epsilon} \left(
\CR+{6 \over R^2} \right) -{1\over 8\pi G} \int_{\partial X_\epsilon} \CK
\right]
.}
The remaining piece of the action computes the  AdS analogue of the ADM mass:
\eqn\admm{
\beta \,M =  \;\lim_{\epsilon \to 0} \left[ -{1\over 16\pi G}
\int_{X-X_\epsilon} \left(\CR+ {6\over R^2}\right) -{1\over 8\pi G}
\int_{\partial (X- X_\epsilon)} \CK
\right]
.}
Of course, the expression for the mass must be normalized by that of the
AdS vacuum. In the process one learns that the boundary term at infinity
gives no contribution. However, the boundary term at the horizon does
give an important positive contribution which makes $M$ grow with $r_0$,
despite the  fact that the bulk action is just measuring the available
volume and this is shrinking.

Then we can normalize the free energy by the volume of the space orthogonal
to the $(r,t)$ plane and define the classical contribution to
the  boundary central charge as: 
\eqn\clab{
\CG_{cl} = {\Delta I \over R^5 \,{\rm Vol} ({\bf S}^5) \,{\rm Vol}({\rm Horizon})}.}
The previous splitting induces two contributions to the central charge:
\eqn\woo{
\CG_{cl} = \CG_{cl}^{\rm cone} + \CG_{cl}^{\rm ADM},}
where the cone contribution is constant,
\eqn\conec{
\CG_{cl}^{\rm cone} = {1\over 8\pi G},}
and the `mass' contribution is monotonically decreasing for $r_0 < R$:
\eqn\massc{
\CG_{cl}^{\rm ADM} = {3 \beta \over 16\pi G} \left({r_0 \over R^2} + {1\over r_0}
\right).}
This is analogous to the boundary central charge in open-string theory (which
 also appears as a classical contribution to the effective action
of open strings.) The  `tension-like' contribution 
to the   boundary central charge decreases as long as the space-time defect
is governed by approximately flat-space  energetics, i.e. for $r_0 <R$.

To conclude, we notice that one can define a notion of `local central
charge' as the integrand of the Euclidean action. This is based on the
fact that, expanding around flat space, a central-charge deficit is
seen in the low-energy effective action as a tree-level 
cosmological constant.  
Thus, if we write
\eqn\rite{
\Delta I = {1\over 20\pi G R^2} \int d{\rm Vol} \;\left[ C(r) -10 \right],}
we can  regard $C(r)$ as a local measure of the number of bosonic
dimensions. In particular $C(r) \rightarrow 10$ 
 as $r\rightarrow \infty$. We find, 
\eqn\loccen{
C(r) = C_\delta \;\delta(r-r_0) + 10\;\Theta(r-r_0),}
where $\Theta$ is the step function and we neglect the cutoff effect
\extras\ that would sligtly increase the asymptotic dimensionality
of space. Thus, we see that the smooth part has $C(r) =0$ for $r<r_0$,
which corresponds to the fact that this part of 
space-time has dissappeared  behind the horizon of the black hole. The
delta-function contribution is proportional to $I_\delta$ and 
negative:
\eqn\deltafff{
C_\delta = -10 \, G\,R^2 \,S  \;\delta.}
 In our previous interpretation of $I_\delta$ we should think
of this term as being smoothed out by short-distance effects in such a way
that the total action is still well approximated by the 
integral \rite. Thus,  
we should have a smooth transition around $r\sim r_0$ between
$C=0$ and $C=10$ spacetimes.

\newsec{Conclusions} 

\noindent

In this note we have developed a physical picture for the decay of
hot type IIB strings in approximately flat ten-dimensional spacetime.
In particular, we have investigated the geometrical interpretation of
the condensation of the Hagedorn tachyon. 

We make use of the AdS/CFT regularization in order to control the 
infrared problems posed by a tachyonic vacuum.      
In this fashion we make the Hagedorn tachyon into a `quasilocalized' tachyon,
similar to those arising at fixed points of nonsupersymmetric orbifolds. 

We propose that the dynamics of the Hagedorn tachyon induces   a 
progressive capping of the thermal manifold in such a way that the
part of space with local temperature above $T_s$ dissappears. 
This picture is natural from various points of view: the geometrical picture
is directly suggested by the Euclidean continuation of the process of
black-hole growth that is expected on the grounds of the string/black-hole
correspondence; it is also 
 suggested by similar situations in the decay of nonsupersymmetric
orbifolds \refs\raps. 

The mechanism for the
 capping could be the some version of the  Kosterlitz--Thouless
RG flow. We notice that, as the effects of the infrared regulator
are felt, i.e. when $r_0 \sim R$, the decay can be described semiclassically
as the growth of a large AdS black hole that is cooler than the
thermal radiation outside it. This continues until the stable AdS black hole
is reached.

If the regulator is removed, the stable Euclidean manifold of the AdS
black hole approaches ${\bf R}^{10}$ with the standard spin structure. 
Therefore, there is a sense in which the string ensemble decays into
a supersymmetric vacuum. It is just the Euclidean geometry of Rindler
space! An immediate consequence of this phenomenon is the conjectured
decay of type 0A unstable vacuum into the type IIB vacuum. Here the  
result has been established for the Wick-rotated theories.

We have also made a preliminary attemp at defining a notion of `boundary
central charge', as in \refs\rchicago. We notice that the natural generalization
of this concept for our set up is as a measure of the effective volume 
left after the geometrical capping.

\vskip0.2cm

\noindent {\bf Acknowledgements}

\vskip 0.1cm

\noindent

We would like to thank L. Alvarez-Gaum\'e and O. Aharony
for useful discussions. The work of E.R. was partially supported by
the Center of excelence project, ISF. American-Israeli bi-National
Science foundation, German-Israel bi-national Science foundation.

\listrefs

\bye